\documentclass[epj]{svjour}

\usepackage{graphicx}
\usepackage{amsmath}
\usepackage{amsfonts}
\usepackage{amssymb}
\usepackage{fancyhdr}
\def\makeheadbox{{%  remove EPJC header
 }}
\def\mytitle{My title} 
\def\myauthors{My name}  
\def\mytype{My type of session}
\def\mysession{My session}
\def\mytitle{Rare K decays in the MSSM} %Put your title here!
\def\myauthors{Christopher Smith}    %Put your name here!
\def\mytype{Contributed Talk}    
\def\mysession{Flavor Physics}
\begin{document}
\title{What rare K decays can tell about the MSSM}
%\subtitle{Do you have a subtitle?\\ If so, write it here}
\author{Christopher Smith\thanks{\emph{Email:}chsmith@itp.unibe.ch}%
}                     % Do not remove
%
%\offprints{}          % Insert a name or remove this line
%
\institute{Institut f\"{u}r Theoretische Physik, Universit\"{a}t Bern, CH-3012 Bern, Switzerland}
%
%\date{Received: date / Revised version: date}
% The correct dates will be entered by Springer
\date{}
\abstract{Supersymmetric contributions to the theoretically clean $K^{+}\rightarrow\pi^{+}\nu\bar{\nu}$,
$K_{L}\rightarrow\pi^{0}\nu\bar{\nu}$, $K_{L}\rightarrow\pi^{0}e^{+}e^{-}$ and $K_{L}\rightarrow\pi^{0}\mu^{+}\mu^{-}$
decays are briefly reviewed. Particular emphasis is laid on the information one could get on the MSSM flavor sector
from a combined study of the four modes.
 % end of PACS codes
} %end of abstract
\maketitle
%DO NOT REMOVE THIS LINE

\section{Introduction}

The FCNC-induced decays, $K^{+}\rightarrow\pi^{+}\nu\bar{\nu}$, $K_{L}%
\rightarrow\pi^{0}\nu\bar{\nu}$, $K_{L}\rightarrow\pi^{0}e^{+}e^{-}$ and
$K_{L}\rightarrow\pi^{0}\mu^{+}\mu^{-}$, are very suppressed in the Standard
Model (SM), where they can be predicted very accurately. Therefore, these
modes are ideal for probing possible New Physics effects\cite{JagerHere}.
In the present talk, the signatures of supersymmetry, in its simplest 
realization as the MSSM, are reviewed.

\section{Rare K decays in the Standard Model}

In the SM, the electroweak processes driving the rare $K$ decays
are the $W$ box, $Z$ and $\gamma$ penguins\cite{BuchallaBL96}, see Fig.1a. In
this section, the excellent theoretical control reached on these contributions
is summarized briefly.\vspace{0.15cm}
\textbf{The $K\rightarrow\pi\nu\bar{\nu}$ decays in the SM}: The
$t$-quark contribution to the Wilson coefficient of the dimension-six FCNC
operator $(\bar{s}d)_{V-A}(\bar{\nu}\nu)_{V-A}$ is known at NLO\cite{BuchallaBL96},
while the $c$-quark one has recently been obtained at NNLO\cite{BurasNNLO}.
The matrix-elements for this operator can be extracted from $K_{\ell3}$ decays,
including NLO isospin corrections\cite{MesciaS07}. For $K^{+}\rightarrow
\pi^{+}\nu\bar{\nu}$, residual $c$-quark effects from dimension-8 operators,
along with long distance $u$-quark contributions, have also been
computed \cite{IsidoriMS}. For $K_{L}\rightarrow\pi^{0}\nu\bar{\nu}$, the
indirect CP-violating contribution (ICPV), $K_{L}\overset{\varepsilon
}{\rightarrow}K_{1}\rightarrow\pi^{0}\nu\bar{\nu}$, is of about 1\%\cite{BB96},
and the CP-conserving one is less than 0.01\%\cite{BI98}.
Altogether, the SM predictions are%
\begin{align*}
\mathcal{B}\left(  K_{L}\rightarrow\pi^{0}\nu\bar{\nu}\right) _{\text{SM}}  &
=(2.49\pm0.39)\cdot10^{-11},\\
\mathcal{B}\left(  K^{+}\rightarrow\pi^{+}\nu\bar{\nu}\right) _{\text{SM}}  &
=(7.83\pm0.82)\cdot10^{-11}.%
\end{align*}
The error on $K_{L}\rightarrow\pi^{0}\nu\bar{\nu}$ is mainly parametric, i.e.
dominated by $\operatorname{Im}\lambda_{t}$, $\lambda_{t}\equiv V_{ts}^{\ast}V_{td}$.
For $K^{+}\rightarrow\pi^{+}\nu\bar{\nu}$, which receives a significant
$c$-quark contribution, the total error could be reduced with a better knowledge
of $m_{c}$ and through a lattice study of higher-dimensional operators\cite{GMT06}.

\textbf{The $K_{L}\rightarrow\pi^{0}\ell^{+}\ell^{-}$ decays in the
SM}: The situation is more involved because there are a priori three
competing processes.

First, the $t$ and $c$-quark contributions, known at NLO\cite{BuchallaBL96},
generate both the dimension-six vector and axial-vector operators:
\[
\mathcal{H}_{eff}=y_{7V}\left(  \bar{s}d\right) _{V}\left(  \bar{\ell}%
\ell\right)  _{V}+y_{7A}\left(  \bar{s}d\right) _{V}\left(  \bar{\ell}%
\ell\right)  _{A}\;.
\]
The former produces the $\ell^{+}\ell^{-}$ pair in a $1^{--}$ state, the
latter in both $1^{++}$ and $0^{-+}$ states.

Secondly, the ICPV contribution is related to $K_{S}\rightarrow\pi^{0}\ell
^{+}\ell^{-}$, which is dominated by the Chiral Perturbation Theory (ChPT)
counterterm $a_{S}$\cite{DambrosioEIP98}. NA48 measurements give $\left|
a_{S}\right|  =1.2\pm0.2$\cite{KSpill}. Producing $\ell^{+}\ell^{-}$ in a
$1^{--}$ state, it interferes with the $(\bar{s}d)_{V}(\bar{\ell}\ell)_{V}$
contribution, arguably constructively\cite{BDI03,FGD04}. This sign could also
be fixed experimentally from $A_{FB}^{\mu}$, the integrated forward-backward,
or muon-energy asymmetry\cite{MesciaST06}.

The final piece is the CP-conserving two-photon contribution, which produces
the lepton pair in either a helicity-suppressed $0^{++}$ or phase-space
suppressed $2^{++}$ state. The LO corresponds to the finite two-loop process
$K_{L}\rightarrow\pi^{0}P^{+}P^{-}\rightarrow\pi^{0}\gamma\gamma\rightarrow
\pi^{0}\ell^{+}\ell^{-}$, $P=\pi,K$, exactly predicted by ChPT, and produces
only $0^{++}$ states. Higher order corrections are estimated using
experimental data on $K_{L}\rightarrow\pi^{0}\gamma\gamma$ for both the
$0^{++}$ and $2^{++}$ contributions\cite{BDI03,ISU04}.

Altogether, the predicted rates are
\begin{align*}
\mathcal{B}\left(  K_{L}\rightarrow\pi^{0}e^{+}e^{-}\right)  _{\mathrm{SM}}  &
=3.54_{-0.85}^{+0.98}\left(  1.56_{-0.49}^{+0.62}\right)  \cdot10^{-11},\\
\mathcal{B}\left(  K_{L}\rightarrow\pi^{0}\mu^{+}\mu^{-}\right)
_{\mathrm{SM}}  & =1.41_{-0.26}^{+0.28}\left(  0.95_{-0.21}^{+0.22}\right)
\cdot10^{-11},%
\end{align*}
for constructive (destructive) interference. The errors are detailed in
\cite{BDI03,MesciaST06,ISU04}, and are currently dominated by the one on
the $K_{S}\rightarrow\pi^{0}\ell^{+}\ell^{-}$ rate measurements.
\begin{figure*}[t]
\centering        \includegraphics[width=0.98\textwidth]{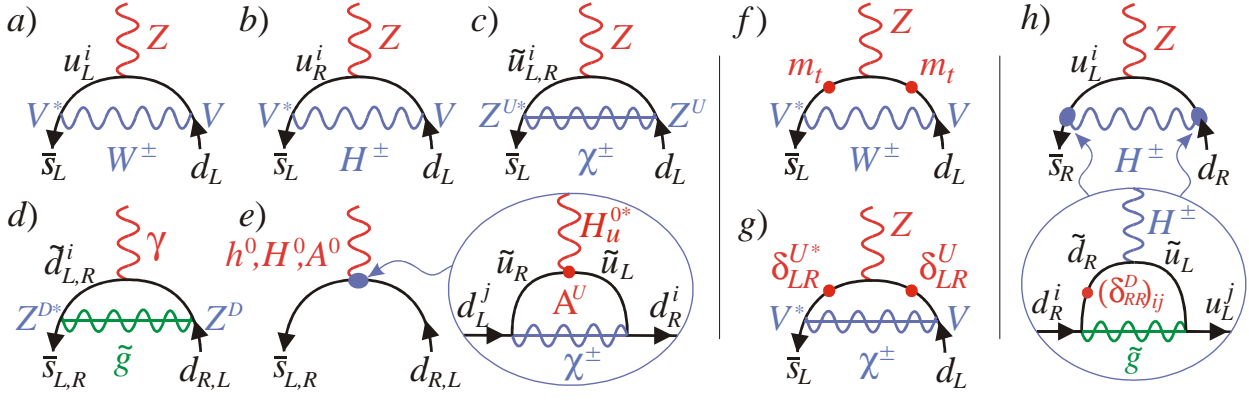}
\caption{$a-e)$ Dominant MSSM contributions to rare $K$ decays. $f-g)$
Dominant sources of $SU(2)_{L}$-breaking in the $Z$-penguin. $h)$ Schematic
representation of the $H^{\pm}$ contribution to the $Z$-penguin at large
$\tan\beta$.}%
\label{fig1}%
\end{figure*}

\section{Rare K decays and supersymmetry}

Even though the minimal supersymmetrization of the SM requires one
super-partner for each SM particle (and two Higgs doublets), it is very
constrained and involves only a few free parameters. However, SUSY must be broken,
and the precise mechanism still eludes us. Therefore, in practice, an effective
description is adopted, introducing all possible explicit soft-breaking terms
allowed by the gauge symmetries. In the squark sector, there are $LL$
and $RR$ mass-terms and trilinear couplings giving rise to $LR$ mass-terms after
the Higgses acquire their VEV's, $\langle H_{u,d}^{0}\rangle=v_{u,d}$:
\begin{align*}
\mathcal{L}_{soft}^{LL,RR} &  =-\tilde{Q}^{\dagger}\mathbf{m}_{Q}^{2}\tilde
{Q}-\tilde{U}\mathbf{m}_{U}^{2}\tilde{U}^{\dagger}-\tilde{D}\mathbf{m}_{D}%
^{2}\tilde{D}^{\dagger}\;,\\
\mathcal{L}_{soft}^{LR} &  =-\tilde{U}\mathbf{A}^{U}\tilde{Q}H_{u}+\tilde
{D}\mathbf{A}^{D}\tilde{Q}H_{d}\;,
\end{align*}
with $\tilde{Q}=(\tilde{u}_{L},\tilde{d}_{L})^{T}$, $\tilde{U}=\tilde{u}_{R}%
^{\dagger}$, $\tilde{D}=\tilde{d}_{R}^{\dagger}$. Obviously, $\mathbf{m}%
_{Q,U,D}^{2}$ and $\mathbf{A}^{U,D}$, which are $\mathbf{3\times3}$ matrices
in flavor-space, generate a very rich flavor-breaking sector as squark mass
eigenstates can differ substantially from their gauge eigenstates.\vspace{0.15cm}

\textbf{What to expect from SUSY in rare $K$ decays:} In the SM, the
$Z$-penguin is the dominant contribution, and is tuned by $\lambda_{t}$ (Fig.1$a$).
The four MSSM corrections depicted in
Figs.1$b-e$ (together with box diagrams), represent the dominant corrections,
and are thus the only MSSM effects for which rare $K$ decays can be sensitive
probes. Let us briefly describe each of them. First, there is the charged
Higgs contribution to the $Z$-penguin (Fig.1$b$), which is, at moderate
$\tan\beta=v_{u}/v_{d}$, aligned with the SM one ($\sim\lambda_{t}$). Then,
there is the supersymmetrized version of Figs.1$a-b$, with charginos --
up-squarks in place of $W^{\pm}/H^{\pm}$ -- up-quarks in the loop (Fig.1$c$),
and which is sensitive to the mixings among the six up-squarks ($Z^{U}$), a
priori not aligned with the CKM mixings. Another purely supersymmetric
contribution, relevant only for charged lepton modes, is the gluino
electromagnetic penguin (Fig.1$d$), sensitive to down-squark mixings ($Z^{D}%
$). The last class of effects consists of neutral Higgs FCNC (Fig.1$e$), and
arises at large $\tan\beta\approx50$. Indeed, the 2HDM-II structure of the
Higgs couplings to quarks, required by SUSY, is not preserved beyond leading
order due to $\mathcal{L}_{soft}$, and the ``wrong Higgs'', $H_{u}$, gets
coupled to down-type quarks, $\mathcal{L}_{eff}\supset\bar{d}_{R}^{i}%
Y_{d}^{ik}(H_{d}^{0}+\epsilon Y_{u}^{\dagger}Y_{u}H_{u}^{0\dagger})^{kj}%
d_{L}^{j}$\cite{HallRS93}. Clearly, once the Higgses acquire their VEV's,
there is a mismatch between quark mass eigenstates and Higgs couplings; both
are no longer diagonalized simultaneously and Higgs FCNC are
gen- erated\cite{BabuKolda99}.\vspace{0.15cm}
\begin{figure*}[t]
\centering        \includegraphics[width=0.98\textwidth]{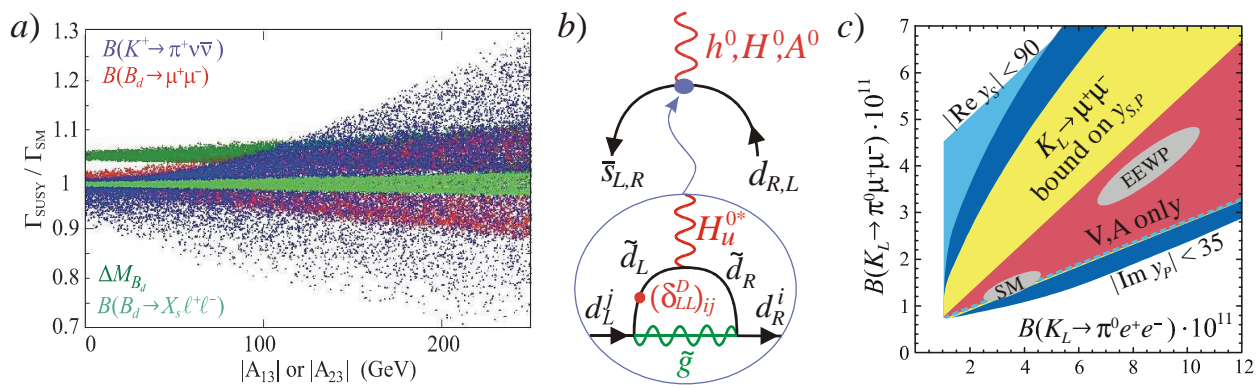}
\vspace{-0.24cm} \caption{$a)$ Sensitivity of $K^{+}\rightarrow\pi^{+}\nu
\bar{\nu}$ to $\mathbf{A}^{U}$ terms, compared to $B$-physics observables.
$b)$ Schematic representation of the neutral Higgs FCNC beyond MFV, at large
$\tan\beta$. $c)$ Impacts of dim-6 FCNC operators in the $\mathcal{B}%
(K_{L}\rightarrow\pi^{0}\mu^{+}\mu^{-})$ vs. $\mathcal{B}(K_{L}\rightarrow
\pi^{0}e^{+}e^{-})$ plane.}%
\label{fig2}%
\end{figure*}

\textbf{Bottom-up approach and Minimal Flavor Violation: }There are
too many parameters in $\mathcal{L}_{soft}$ to have any hope to fix them all
from rare $K$ decays. At the same time, however, the observed suppression of
FCNC transitions and CP-violating phenomena seem to indicate that only small
departures with respect to the SM are possible. Therefore, one starts from a
lowest-order basis in which the flavor-breakings due to $\mathbf{m}_{Q,U,D}^{2}$
and $\mathbf{A}^{U,D}$ are minimal. This can take the form of $mSUGRA$, 
alignment of squarks with quarks or the Minimal Flavor Violation hypothesis (MFV).
In a second stage, one probes the possible signatures of departures from this minimal
setting. The goal being, ultimately, to constrain SUSY-breaking models, which
imply specific soft-breaking structures. At that stage, information from rare
$K$ decays, colliders and $B$-physics must of course be combined.

Here we adopt MFV as the lowest order basis, i.e. we impose that the SM
Yukawas $\mathbf{Y}_{u,d}$ are the only sources of flavor-breaking\cite{MFV}.
In practice, this means that $\mathcal{L}_{soft}$ terms can be expanded as
($a_{i},b_{i}\sim O(1)$, and $A_{0}$, $m_{0}$ set the supersymmetry-breaking scale)%
\begin{align*}
\mathbf{m}_{Q}^{2} &  =m_{0}^{2}(a_{1}\mathbf{1}+b_{1}\mathbf{Y}_{u}^{\dagger
}\mathbf{Y}_{u}+b_{2}\mathbf{Y}_{d}^{\dagger}\mathbf{Y}_{d}\\
&  \;\;\;\;+b_{3}(\mathbf{Y}_{d}^{\dagger}\mathbf{Y}_{d}\mathbf{Y}_{u}^{\dagger
}\mathbf{Y}_{u}+\mathbf{Y}_{u}^{\dagger}\mathbf{Y}_{u}\mathbf{Y}_{d}^{\dagger
}\mathbf{Y}_{d})),\\
\mathbf{m}_{U}^{2} &  =m_{0}^{2}(a_{2}\mathbf{1}+b_{4}\mathbf{Y}_{u}%
\mathbf{Y}_{u}^{\dagger}),\\
\mathbf{A}^{U} &  =A_{0}\mathbf{Y}_{u}(a_{4}\mathbf{1}+b_{6}\mathbf{Y}%
_{d}^{\dagger}\mathbf{Y}_{d})\;,
\end{align*}
and similarly for $\mathbf{m}_{D}^{2}$ and $\mathbf{A}^{D}$, such that all FCNC's
and CP-violation are still essentially tuned by the CKM matrix.
For example, the dominant contributions to the $Z$-penguin are those
breaking the $SU(2)_{L}$ gauge-symmetry\cite{NirWorah98,BurasRS98}. In the SM,
this breaking is achieved through a double top-quark mass insertion
(Fig.1$f$). Similarly, in the MSSM, it is the double $\tilde{t}_{L}-\tilde
{t}_{R}$ mixing via the $\mathbf{A}^{U}$ trilinear terms which plays the
dominant role (Fig.1$g$ in the sCKM basis)\cite{ColangeloI98}. Within MFV,
this gives a factor $m_{t}^{2}$ $\lambda_{t}$ $\left|  a_{4}-\cot\beta
\mu^{\ast}\right|  ^{2}$ \cite{IsidoriMPST06}, still enhanced by $m_{t}^{2}$
and tuned by $\lambda_{t}$.

\section{Supersymmetric effects in $K\rightarrow\pi\nu\bar{\nu}$}

\textbf{SUSY effects in the (axial-)vector operators}, $(\bar
{s}d)_{V\pm A}(\bar{\nu}\nu)_{V-A}$, cannot be distinguished since only
$(\bar{s}d)_{V}(\bar{\nu}\nu)_{V-A}$ contributes to the $K\rightarrow\pi
\nu\bar{\nu}$ matrix-element. All MSSM effects are thus encoded into a single
complex number, $X^{\nu}\equiv y_{L}^{\nu}+y_{R}^{\nu}$ \cite{BurasRS98}:
\begin{align*}
\mathcal{H}_{eff}  &  =y_{L}^{\nu}\left(  \bar{s}d\right)  _{V-A}\left(
\bar{\nu}\nu\right)  _{V-A}+y_{R}^{\nu}\left(  \bar{s}d\right)  _{V+A}\left(
\bar{\nu}\nu\right)  _{V-A}\\
&  \rightarrow\left(  y_{L}^{\nu}+y_{R}^{\nu}\right)  \left(  \bar{s}d\right)
_{V}\left(  \bar{\nu}\nu\right)  _{V-A}\;.
\end{align*}
At moderate $\tan\beta$, chargino penguins are the dominant MSSM contributions
because of their quadratic sensitivity to up-squark mass-insertions
(Figs.1$c$, 1$g$). Within MFV, this means, given the $m_{t}$ enhancement
present in the $\delta_{LR}^{U}$ sector, that $K\rightarrow\pi\nu\bar{\nu}$
are particularly sensitive. Still, a significant enhancement would require a
very light stop and chargino\cite{IsidoriMPST06}, mostly because of the
constraint from $\Delta\rho$\cite{BurasGGJS00}. Any enhancement $\gtrsim5\%$
would thus falsify MFV if sparticles are found above $\sim200GeV$, and if
$\tan\beta\gtrsim5$ (to get rid of the $H^{\pm}$ contribution). Turning on
generic $\mathbf{A}^{U}$ terms, the largest deviations arise in $K\rightarrow
\pi\nu\bar{\nu}$, see Fig.2$a$\cite{IsidoriMPST06}. Further, the decoupling is
slower than for observables sensitive to chargino boxes like $\varepsilon_{K}%
$. All in all, given that $K^{+}\rightarrow\pi^{+}\nu\bar{\nu}$ has already
been seen, how large the effect could be for $K_{L}\rightarrow\pi^{0}\nu
\bar{\nu}$? By an extensive, adaptive scanning over the MSSM parameter space,
Ref.\cite{BurasEJR05} has shown that the GN model-independent bound\cite{GrossmanN97}
can be saturated, which represents a factor $\sim30$ enhancement of
$\mathcal{B}(K_{L}\rightarrow\pi^{0}\nu\bar{\nu})$ over the SM.

At large $\tan\beta$, the chargino contributions may no longer represent the
dominant effect. While the Higgs FCNC obviously does not contribute
(Fig.1$e$), higher order effects in the $H^{\pm}$ contribution to the
$Z$-penguin (Fig.1$h$), sensitive to $\delta_{RR}^{D}$, can become sizeable
beyond MFV\cite{IsidoriP06}. Further, this contribution is slowly decoupling
as $M_{H}$ increases compared to tree-level neutral Higgs exchanges, as for
example in $B_{s,d}\rightarrow\mu^{+}\mu^{-}$.\vspace{0.15cm}

\begin{table*}[ptb]
\caption{Sensitivity of rare $K$ decays to MSSM effects, with and without
MFV, and with moderate and large $\tan\beta$. The dominant
contributions come from single, $(\delta_{j}^{i})_{12}$, and/or double
(e.g. $(\delta_{j}^{i})_{32}^{\ast}(\delta_{j}^{i})_{31}$) mass insertions (see text).}%
\label{TableCCL}
\centering   {
\begin{tabular}
[c]{|c|c|c|}\hline
MSSM scenario & $K\rightarrow\pi\nu\bar{\nu}$ & $K_{L}\rightarrow\pi^{0}%
\ell^{+}\ell^{-}$\\\hline
MFV, $\tan\beta\approx2$ & Best sensitivity, but maximal & Less sensitive, but
precisely\\
& enhancement $<$ 20-25\% & correlated with $K\rightarrow\pi\nu\bar{\nu}%
$\\\hline
MFV, $\tan\beta\approx50$ & \multicolumn{2}{|c|}{Negligible effects ?}\\\hline
General, $\tan\beta\approx2$ & Best probes of $\delta_{LR}^{U}$ &
\multicolumn{1}{|l|}{$\delta_{LR}^{U}:$ correlated with $K\rightarrow\pi
\nu\bar{\nu}$}\\
& (quadratic dependence in $\delta_{LR}^{U}$) & \multicolumn{1}{|l|}{$\delta
_{LR}^{D}:$ correlated with $\varepsilon^{\prime}/\varepsilon$ (but
cleaner)}\\\hline
General, $\tan\beta\approx50$ & Good probes of $\delta_{RR}^{D}$ & Good probes
of $\delta_{RR,LL}^{D}$,\\
& (slow decoupling as $M_{H} \rightarrow\infty$) & correlated with $K_{L}%
\rightarrow\mu^{+}\mu^{-}$ (but cleaner)\\\hline
\end{tabular}
}\end{table*}

\textbf{SUSY effects in other dimension-six operators}, $(\bar{s}d)(\bar{\nu
}(\mathbf{1},\gamma_{5})\nu)$ and $(\bar{s}\sigma_{\mu\nu}d)(\bar{\nu}%
\sigma^{\mu\nu}(\mathbf{1},\gamma_{5})\nu)$, require active right-handed
neutrinos and will not be discussed here\cite{OtherOperators}. Another
possible class of operators, since the neutrino flavors are not detected, are
$(\bar{s}\Gamma^{A}d)(\bar{\nu}^{i}\Gamma^{B}\nu^{j})$ with $i\neq j$ and
$\Gamma^{A,B}$ some Dirac structures. In the MSSM, such lepton-flavor
violating operators arise only from suppressed box diagrams, and cannot lead
to significant effects\cite{GrossmanIM04}. However, they could be sizeable in
the presence of R-parity violating terms\cite{GrossmanIM04,Rparity1}.

\section{Supersymmetric effects in $K_{L}\rightarrow\pi^{0}\ell^{+}\ell^{-}$}

Though the SM predictions for these modes are less accurate than for
$K\rightarrow\pi\nu\bar{\nu}$, they are sensitive to more types of New Physics
operators\cite{MesciaST06}. Indeed, the final-state leptons are now charged
and massive. Therefore, besides electromagnetic effects, common to both the
muon and electron modes, the relatively large muon mass opens the possibility
to probe a whole class of helicity-suppressed effects.\vspace{0.15cm}

\textbf{SUSY effects in the QCD operators}, i.e. in the chromomagnetic
$\bar{s}\sigma_{\mu\nu}dG^{\mu\nu}$ or four-quark operators, have no direct
impact on $K_{L}\rightarrow\pi^{0}\ell^{+}\ell^{-}$. Indeed, as said in
Sect. 2, the two-photon CPC piece is fixed entirely in terms of the
measured $K\rightarrow\pi\pi\pi$, $\pi\gamma\gamma$ modes\cite{BDI03,ISU04},
while the ICPV contribution is fixed from the measured $\varepsilon_{K}$ and
$K_{S}\rightarrow\pi^{0}\ell^{+}\ell^{-}$ rate\cite{DambrosioEIP98}. At the
low scale $\mu\lesssim m_{c}$, new physics can thus explicitly enter
through semi-leptonic FCNC operators only.\vspace{0.15cm}

\textbf{SUSY effects in the SM operators}, which are the vector and
axial-vector operators, can in principle be disentangled thanks to the different
sensitivities of the two modes to the axial-vector current (as discussed in
Sec. 2, it also produces $\ell^{+}\ell^{-}$ in a helicity-suppressed
$0^{-+}$ state). Various MSSM contributions can enter in $y_{7A}$ and $y_{7V}$.
First, chargino contributions to the $Z$-penguin (Fig.1$c$) enter as
$y_{7A},y_{7V}\sim(\delta_{RL}^{U})_{32}^{\ast}(\delta_{RL}^{U})_{31}$,
and are thus directly correlated to the corresponding contribution to
$K\rightarrow\pi\nu\bar{\nu}$\cite{IsidoriMPST06,ChoMW96}. Within MFV, the
maximal effect for $K_{L}\rightarrow\pi^{0}\ell^{+}\ell^{-}$ is about one third
of the one for $K_{L}\rightarrow\pi^{0}\nu\bar{\nu}$, hence may be inaccessible
due to theoretical uncertainties. Secondly, gluino contributions to the
electromagnetic operator $\bar{s}\sigma_{\mu\nu}dF^{\mu\nu}$ (Fig.1$d$) can be
absorbed into $y_{7V}\sim(\delta_{RL}^{D})_{12}$. Even if directly correlated
with $\varepsilon^{\prime}/\varepsilon$, sizeable effects in $K_{L}%
\rightarrow\pi^{0}\ell^{+}\ell^{-}$ are still possible\cite{BurasCIRS99}.
Finally, $H^{\pm}$ contributions arise at large $\tan\beta$ (Fig.1$h$), with
$y_{7A},y_{7V}\sim(\delta_{RR}^{D})_{12}$, and are directly correlated with
those for $K\rightarrow\pi\nu\bar{\nu}$\cite{IsidoriP06}.\vspace{0.15cm}

\textbf{SUSY effects in the (pseudo-)scalar operators}, which can be
helicity-suppressed or not:
\begin{align*}
\mathcal{H}_{eff}  &  = y_{S}\left(  \bar{s}d\right)  \left(  \bar{\ell}%
\ell\right)  +y_{P}\left(  \bar{s}d\right)  \left(  \bar{\ell}\gamma_{5}%
\ell\right) \\
&  \;\; +y_{S}^{\prime}\left(  \bar{s}\gamma_{5}d\right)  \left(  \bar{\ell
}\ell\right)  +y_{P}^{\prime}\left(  \bar{s}\gamma_{5}d\right)  \left(
\bar{\ell}\gamma_{5}\ell\right)  \;.
\end{align*}
The first (last) two operators contribute to $K_{L}\rightarrow\pi^{0}\ell
^{+}\ell^{-}$ ($K_{L}\rightarrow\ell^{+}\ell^{-}$). In the MSSM at large
$\tan\beta$, they arise from Higgs FCNC\cite{IsidoriRetico}, and are thus
helicity-suppressed (Fig.2$b$). Sizeable effects for the muon mode are
possible beyond MFV, where they are sensitive to $(\delta_{RR,LL}^{D})_{12}$
and $(\delta_{RR}^{D})_{23}(\delta_{LL}^{D})_{31}$ mass-insertions. Also, even
if this contribution is correlated to the one for $K_{L}\rightarrow\mu^{+}%
\mu^{-}$, given the large theoretical uncertainties for this mode, a factor
$\sim4$ enhancement is still allowed (Fig.2$c$)\cite{MesciaST06}. On the other
hand, helicity-allowed contributions to these operators do not arise in the
MSSM. They could appear in the presence of R-parity violating couplings, but,
baring fine-tuning, their effects must be small to avoid overproducing
$K_{L}\rightarrow e^{+}e^{-}$\cite{MesciaST06}.\vspace{0.15cm}

\textbf{SUSY effects in the (pseudo-)tensor operators}, $(\bar{s}\sigma
_{\mu\nu}d)(\bar{\ell}\sigma^{\mu\nu}(\mathbf{1},\gamma_{5})\ell)$, the last
possible dimension six semi-leptonic FCNC operators, are helicity-suppressed
in the MSSM\cite{BobethBKU02} and, being also phase-space suppressed, do not
lead to any significant effect \cite{MesciaST06}. Further, they cannot arise
from $R$-parity violating couplings.

\section{Conclusion}

The $K^{+}\rightarrow\pi^{+}\nu\bar{\nu}$,
$K_{L}\rightarrow\pi^{0}\nu\bar{\nu}$, $K_{L}\rightarrow\pi^{0}e^{+}e^{-}$ and
$K_{L}\rightarrow\pi^{0}\mu^{+}\mu^{-}$ decay modes are the only theoretically
clean windows into the $\Delta S=1$ sector. If SUSY is discovered, the pattern of
deviations they could exhibit with respect to the SM (see Table 1) will be
essential to constrain the MSSM parameter-space, and hopefully unveil the
nature of the SUSY-breaking mechanism.

\end{document}